\documentclass[a4paper,11pt]{article}
\pdfoutput=1 

\usepackage{jcappub} 

\usepackage[T1]{fontenc} 
\usepackage{footnote}

\title{\boldmath{Constraints on cosmological birefringence energy dependence from CMB polarization data}}

\author[a,b]{G. Gubitosi}
\author[c,d]{and F. Paci}

\affiliation[a]{Dipartimento di Fisica and sez. Roma1 INFN, Universit\`a di Roma ``La Sapienza'',\\P.le A. Moro 2, 00185 Rome, Italy}
\affiliation[b]{CEA, IPhT,\\91191 Gif-sur-Yvette cedex, France}
\affiliation[c]{SISSA, Scuola Internazionale Superiore di Studi Avanzati,\\Via Bonomea 265, Trieste, 34136, Italy}
\affiliation[d]{APC, Univ. Paris Diderot, CNRS/IN2P3, CEA/Irfu, Obs. de Paris, Sorbonne Paris Cit\'e,  \\10, rue A. Domon et L. Duquet, 75205 Paris cedex 13, France}

\emailAdd{giulia.gubitosi@roma1.infn.it}
\emailAdd{fpaci@sissa.it}

\abstract{
We study the possibility of constraining the energy dependence of cosmological birefringence by using CMB polarization data. We consider four possible behaviors, characteristic of different theoretical scenarios: energy-independent birefringence motivated by Chern-Simons interactions of the electromagnetic field, linear energy dependence motivated by a `Weyl' interaction of the electromagnetic field, quadratic energy dependence, motivated by quantum gravity modifications of low-energy electrodynamics, and inverse quadratic dependence, motivated by Faraday rotation generated by primordial magnetic fields.
We constrain the parameters associated to each kind of dependence and use our results to give constraints on the models mentioned.  We forecast the sensitivity that Planck data will be able to achieve in this respect.
}

\begin{document}
\maketitle
\flushbottom

\section{Introduction}
\label{sec:intro}

The improvement of experimental techniques for measuring the Cosmic Microwave Background radiation (CMB) has allowed  to measure with increasing accuracy the parameters of the standard cosmological model \cite{Komatsu:2010fb}. 
In particular, a boost to this field was given by the ability to not only measure the CMB temperature anisotropies, but also its polarization, whose underlying physics is now very well understood, and which can give valuable information on crucial aspects of the universe history, such as the reionization epoch and the inflationary mechanism.

The precision achievable with the last generation of experiments allows us to be ambitious and start considering CMB data as a tool to test also models of non-standard physics. 
In particular, it is possible to constrain cosmological birefringence effects, and since the first  WMAP and BOOMERanG data were analyzed with this aim \cite{Feng:2006dp}, tens of papers appeared showing how very significant constraints can be already put on this kind of effect.

Birefringence produces a rotation of the polarization direction of a linearly polarized electromagnetic (EM) field during its propagation. For CMB radiation, this causes a peculiar mixing between the E and B modes of the polarization field \cite{Lue:1998mq, Lepora:1998ix}.
In particular, if we indicate with $\alpha$ the amount of rotation, the effect on CMB angular power spectra can be formalized as:
\begin{eqnarray}
 C_\ell^{EE}&=&\tilde C_\ell^{EE}  \cos^2\left(2 \alpha\right) +\tilde C_\ell^{BB}  \sin^2\left(2 \alpha\right)  \nonumber \\
C_\ell^{BB}&=&\tilde C_\ell^{EE}  \sin^2\left(2 \alpha\right)  +\tilde C_\ell^{BB}  \cos^2\left(2 \alpha\right) \nonumber \\
C_\ell^{EB}&=&\frac{1}{2}\left(\tilde C_\ell^{EE}-\tilde C_\ell^{BB}\right)  \sin\left(4 \alpha\right)     \nonumber \\
C_\ell^{TE}&=&\tilde C_\ell^{TE}  \cos\left(2 \alpha\right)     \nonumber \\
C_\ell^{TB}&=&\tilde C_\ell^{TE}  \sin\left(2 \alpha\right)     \label{eq:powerspectra}
\end{eqnarray}
where $\tilde C_\ell$ refer to the spectra that one would expect without any birefringence effect, and $C_\ell$ are the measured spectra if a polarization rotation has indeed happened.
Note that in particular birefringence predicts non zero TB and EB cross-correlation, which would be instead vanishing according to standard parity-preserving physics.

Motivations for testing cosmological birefringence arise in many contexts of theoretical physics.
In general a mixing between E and B modes, which have opposite properties under parity transformations, signals the occurrence of some parity-violating phenomenon, but the details of the mechanism can be very different.

Already standard physics predicts a mixing between E and B modes as described in eq.(\ref{eq:powerspectra}), if CMB photons happen to travel through a magnetic field. This effect is known as  Faraday rotation, and can affect the whole CMB sky if caused by primordial magnetic fields \cite{Kosowsky:1996yc, Giovannini:1997km}. 
Besides this, there are several non-standard physics models implying birefringent behavior of light, which are well motivated from a theoretical perspective. Some examples are  the coupling of the electromagnetic (EM) field with a quintessence field \cite{Giovannini:2004pf, Balaji:2003sw, Liu:2006uh}, coupling of the EM field with axions \cite{Finelli:2008jv}, Chern-Simons interactions in the EM Lagrangian \cite{Carroll:1989vb, Xia:2008si, Xia:2009ah}  and quantum-gravity motivated effective theories for electromagnetism, introducing non-renormalizable, parity-violating, corrections to the EM Lagrangian \cite{Myers:2003fd, Kahniashvili:2008va, Gubitosi:2009eu}.

This plethora of different models calls for finding a way to discriminate between them if any birefringence signature in CMB polarization data appears.
We will  focus on the main feature that could make this possible, at least in principle, i.e. the energy dependence of the induced polarization rotation.

Presently available constraints, obtained by combining different CMB datasets, assume energy-independent birefringence (the most recent and complete analysis is \cite{Xia:2012ck}).
This can provide general information on the strength of the birefringence effect, but is of course very limiting for testing specific models predicting non-trivial energy dependencies.

We will show that, although the quality of current CMB data is not yet good enough to discriminate among the energy dependencies investigated here, constraints on the relevant theoretical models may be significantly improved by properly accounting for the observational frequency bands. 
We will also show that a great improvement  is to be expected once Planck data will be analyzed in the way we suggest.

The next section provides a review of current constraints on birefringence coming from the analysis of WMAP7, BOOMERanG, QUAD and BICEP data.  We also give a review on the combined analyses proposed so far.
In section III we review non-standard models which predict specific energy dependence of the birefringence effect. We then propose a method to combine datasets (different experiments and difference frequency channels) according to the energy dependencies we want to test and we provide our constraints.
In section IV we forecast on the sensitivity reachable by the Planck mission, if an appropriate treatment of the information coming from different frequency channels is performed. 
We discuss Planck capabilities of distinguishing among different energy dependencies and show that the expected sensitivity on energy-dependent effects improves up to
one order of magnitude with respect to what it is achievable with CMB present data.
Section V provides a discussion of our results in light of the different models we have tested, and translates our results into constraints on the specific model parameters. We show that for some of the models  our method allows to improve present constraints of about one order of magnitude.

\section{Current constraints}

In this section we briefly report and discuss present available constraints on birefringence coming from CMB polarization observations. 

Since our main goal is to provide an energy dependence  analysis, ideally we would only report results coming from the analysis of single energy channels. 
However, because not all the constraints available in the literature are of this kind, we will also consider those drawn from multifrequency experiments, where the different channels have been combined. The issue of  accounting for the frequency dependence implicitly encoded in such results will be discussed in the next section. 

The most recent constraints on $\alpha$ from CMB observation  are shown in table \ref{tab:constraints}, which also shows  the frequencies observed, the multipole range explored and the uncertainty associated to the estimates. 

\begin{savenotes}

\begin{table}[htdp]
\centering

\begin{tabular}{|c|c|c|c|c|}
\hline
Experiment&Energy channels  (GHz)&$\ell$&$\alpha\pm stat(\pm syst)$ (deg)&Reference\\
\hline
WMAP7&33+41+61&$2-23$&$-3.8\pm5.2\pm 1.5$&\cite{Komatsu:2010fb}\\
{\bf WMAP7}&{\bf 41+61+94}&{\bf 24-800 }&$\mathbf{-0.9\pm1.4\pm1.5}$&$\textbf{\cite{Komatsu:2010fb}}$\\
WMAP7&33+41+61+94 \footnote{The lower (higher) frequency does not contribute to the high (low) multipole constraint (cfr. the first two lines in the table)}&$2-800$&$-1.1\pm1.4\pm 1.5$&\cite{Komatsu:2010fb}\\
WMAP7&33+41+61&$2-23$&$-3.0^{+2.6}_{-2.5}\;$  \footnote{For this result and the following one should add the same systematic as for other WMAP7 results}&\cite{Gruppuso:2011ci}\\
WMAP7&33+41+61&$2-47$&$-1.6\pm1.7$&\cite{Gruppuso:2011ci}\\
WMAP7&33+41+61&$2-30$&$-4.2^{+1.9 \, +10.2}_{-3.1 -7.5}$&\cite{Finelli:2012wu}\\
WMAP7&33+41+61&$2-800$&$-1.3^{+0.6 +2.3}_{-0.7 -2.3}$&\cite{Finelli:2012wu}\\
{\bf BOOM03}&{\bf145}&{\bf 150-1000}&$\mathbf{-4.3\pm 4.1}$ \footnote{This error already takes into account BOOMERanG systematic error of $-0.9^{\circ}\pm0.7^{\circ}$}&\textbf{\cite{systematics}}\\
{\bf QUAD}&{\bf 100}&{\bf 200-2000}&$\mathbf{-1.89\pm2.24\pm 0.5}$&\textbf{\cite{Wu:2008qb}}\\
{\bf QUAD}&{\bf 150}&{\bf 200-2000}&$\mathbf{0.83\pm0.94\pm 0.5}$&\textbf{\cite{Wu:2008qb}}\\
QUAD&100+150&$200-2000$&$0.64\pm 0.5\pm0.5$&\cite{Brown:2009uy}\\
{\bf BICEP}&{\bf 100+150}&{\bf 21-335}&$\mathbf{-2.60\pm1.02\pm 0.7}$ &\textbf{\cite{Xia:2009ah}}\footnote{The systematic uncertainty is given in \cite{Takahashi:2009vp}}\\
\hline
\end{tabular}
\caption{Presently available constraints on birefringence angle $\alpha$ from CMB. We report the experiment from which data were taken, the observational frequency channels, the harmonic multipole range explored, the estimated value of the birefringence angle, together with its statistical  and  systematic uncertainty (unless differently stated in the footnotes). The lines in boldface correspond to the data that we actually use for our analysis (see the text for further details).}
\label{tab:constraints}
\end{table}%

\end{savenotes}

For QUAD data we report both the combined-channel result and the constraints coming from the two single-channel analysis. The former results from an updated analysis, and gives in principle stronger constraints. Unfortunately, no update on the single frequency results has been reported by the QUAD team. Since we are mostly interested in analyzing potential energy dependence of the birefringence effect, we will in the following work with the QUAD single channel results, even if they are less recent.

Constraints on $\alpha$ from WMAP7 data have been put either on the whole range of multipoles accessible, or by focusing on the low and high multipoles separately. Multipoles below $\ell\sim 20$ are expected to be mostly sensitive to photons that underwent reionization, at $z\sim 10$. Since reionization resets the polarization anisotropies produced at the last scattering surface, the observed birefringence effect would result as a superposition of contributions belonging to different scales and times. 
In order to make our dataset uniform and concentrate only on the energy dependence of birefringence, we will include in our study only results obtained for the multipole range $24\div 800$.

The WMAP team also reports on a combined constraint on the birefringence angle from WMAP7, QUAD and BICEP, where an inverse variance weighting scheme is adopted and systematic errors are taken into account by adding them (in quadrature) to the statistical ones \cite{Komatsu:2010fb}. They find  $\alpha=-0.25^{\circ}\pm0.58^{\circ}$, which is well compatible with no polarization rotation effect. 
In \cite{Xia:2012ck}  the constraint from WMAP7, BOOMERanG and BICEP observations is  $\alpha=-2.28^{\circ}\pm1.02^{\circ}$, which results from a joint analysis of all angular power spectra, and takes into account also systematic uncertainties. This gives a $2.2$ standard deviations indication for a birefringence effect. However, such indication is strongly reduced by the addition of QUAD data, providing $\alpha=-0.26^{\circ}\pm0.54^{\circ}$, including systematic uncertainties.

This is of course due to the fact that the estimate of the polarization rotation angle coming from QUAD data (combined energy channels) has some tension with results from other experiments (a $\sim 2\sigma$ tension under the assumption of energy-independent effect \cite{Xia:2012ck}), which comes mostly from the $150$ GHz channel (see table \ref{tab:constraints}). A brief discussion about such tension is  provided in \cite{Xia:2012ck} and \cite{Xia:2009ah}\footnote{Note that in \cite{Xia:2012ck} also constraints from the single experiments are re-derived, both including and not including systematic errors. 
These  are in very good agreement with the ones reported in our table \ref{tab:constraints}, once statistical and systematic errors are simply summed in quadrature.}. We will come back on the topic in the next section.

All the  constraints discussed above, obtained combining data from different experiments, assume no energy dependence of the birefringence rotation angle. This is of course limiting if one wants to put reliable constraints on an effect which might very well have some non trivial dependence on the radiation energy. In the next Section we will focus on how to exploit current estimates of $\alpha$ to investigate its energy dependence.

\section{Study of possible energy dependence of birefringence effects }
In this section we  test the hypothesis that birefringence of CMB photons might have some specific energy dependence. As we mentioned in the introduction, this can be a powerful method to discriminate among different theoretical models predicting a rotation of polarization during  photons propagation. 
Here we will limit ourselves to qualitatively describe the dependence of the polarization rotation angle $\alpha$ as a function of photon energy $E$. In the final discussion section we will make actual connection between these qualitative behaviors and specific models, relating the parameters used in this and the following section with the parameters characterizing the models considered.\\
The laws of energy dependence that we will test are of four kinds:

\begin{itemize}
\item square law dependence ($\alpha (E)= q E^{2}$). This is predicted by quantum gravity-inspired models, such as the Lorentz violating effective field theory for Electrodynamics proposed in \cite{Myers:2003fd} and was already considered in a cosmological framework in \cite{Gubitosi:2009eu, Kahniashvili:2008va}. 
\item linear dependence ($\alpha (E)= l E$). This kind of dependence is motivated by  quintessence coupling models (eventually string-inspired) \cite{Balaji:2003sw}, and more in general by Lorentz violating electrodynamics, such as renormalizable 
`Weyl' interactions \cite{Shore:2004sh,Kahniashvili:2008va}. 
\item no dependence ($\alpha(E)= \alpha_{0}$). This is the most studied case in the CMB literature, and basically the only one considered when different datasets are combined (with a few exceptions, such as a work by one of the authors \cite{Gubitosi:2009eu}) 
Energy independent birefringence is expected in the presence of Chern-Simons interactions of the EM field \cite{Carroll:1989vb, Shore:2004sh}.
\item inverse square law dependence ($\alpha(E)= h E^{-2}$). This is the kind of dependence expected from  Faraday rotation due to the photons traveling through a magnetic field. In particular a uniform rotation across the sky can be traced back to the presence of primordial homogeneous magnetic fields.
\end{itemize} 

The parameters $q$, $l$, $\alpha_{0}$, $h$ are just phenomenological parameters related to the different energy dependencies. In the discussion section we link them to actual theoretical models parameters.

\subsection{Combining different datasets}
Our approach will be to fit for the four kinds of energy dependence described above using the available constrains on birefringence angle coming from individual CMB experiments, reported in boldface in table \ref{tab:constraints}.

In combining the constraints coming from different experiments in such a way that allows for a study of the energy dependence we have to face two issues.
The first one is that  each energy channel to which a CMB experiment is sensitive will have a certain width.
Instrumental response per frequency is usually very well modeled by a constant function within a given frequency range, its bandwidth, which is of order $20\div 30\%$ of the nominal energy.  In table \ref{tab:bandwidths} we report the frequency channels  used by the various experiments  to constrain birefringence and the channels bandwidth.
The second issue is the fact that for some experiments (WMAP7 and BICEP, see table \ref{tab:constraints}) the  constraints are derived from a combination of several channels,  under the assumption that the amount of polarization rotation does not depend on energy.

\begin{table}[htdp]
\centering
\begin{tabular}{|c|c|c|}
\hline
Experiment&Energy  (GHz)&Bandwidth (GHz)\\
\hline
WMAP7&41&8\\
&61&14\\
&94&20\\
\hline
BOOM03&145&45\\
\hline
QUAD&95&26\\
\hline
QUAD&150&41\\
\hline
BICEP&96&22\\
&150&39\\
\hline
\end{tabular}
\caption{Energy channels and corresponding bandwidth of the experiments considered in the present work.}
\label{tab:bandwidths}
\end{table}

Let us formalize the issue of properly accounting for multifrequency and bandwidths.
If the true functional dependence of $\alpha$ on energy is $\alpha =\alpha(E)$, then we can interpret the measured value $\alpha_{f}$ from each experiment as  
the result
we would have obtained if we had only considered data at some effective energy $E_{f}$, i.e. $\alpha_{f}\equiv \alpha(E_{f})$. Of course the actual value of $E_{f}$ depends on the specific functional dependence of $\alpha$ on $E$.
The relation between $\alpha_{f}$ and the function $\alpha(E)$ can be written as:
\begin{equation}
\alpha_f=\frac{1}{N}\int \alpha(E) f(E)dE\,,\qquad N=\int f(E)d E
\label{alpha_f}
\end{equation}
where $f(E)$ is a weighting function.
 We assume a constant response for each frequency channel $i$ within a given energy interval $[E^{\,(i)}_-,E^{\,(i)}_+]$ (its bandwidth) and
we will adopt a weighting scheme of the different channels according to the corresponding pixel nominal sensitivity \footnote{This is coherent with the assumption, made in \cite{Komatsu:2010fb},  that the uncertainty on $\alpha$ is related to the signal to noise ratio of TE and EE in the following way: $Err[\alpha^{TB} ] \simeq	\frac{1}{2(S/N )^{TE}},\;
Err[\alpha^{EB}]\simeq \frac{1}{2(S/N )^{EE}}$.}.
 So:
\begin{equation} 
f(E) = \left\{ \begin{array}{ll}
1/\sigma_i^2  & \quad \textrm{if $ E \in [E^{\,(i)}_{-},E^{\,(i)}_+]$}\\ 
0 & \quad \textrm{elsewhere}
\end{array} \right. 
\label{fofE}
\end{equation}
where $\sigma_i$ is the instrumental noise per pixel. From Eqs. (\ref{alpha_f}) and (\ref{fofE}), the effective energy $E_f$ associated to an experiment to can be computed for any kind of assumed energy dependence.

For clearness, we provide in Appendix \ref{sec:effectiveEnergy} the explicit example of a multi-channel experiment (as WMAP and BICEP for the purposes of the present work), for the different kinds of energy dependence under investigation.

By means of the procedure sketched above, for each functional energy dependence we want to test we translate the information of table \ref{tab:bandwidths} into an effective energy $E_{f}$ to be associated to the  value of $\alpha$ measured by each experiment. Results are reported in table \ref{tab:energychannels}, together with the corresponding constraint on $\alpha$ 
and the associated error. The latter includes statistical and systematic uncertainties, taken as independent and so summed in quadrature. 

To test a specific energy dependence, we simply perform a best-fit analysis of the data in table \ref{tab:energychannels} to constrain the parameter of the models, $\alpha_0, l, q, h$.

\begin{table}[htdp]
\centering
\begin{tabular}{|c|c|c|c|}
\hline
Experiment&Energy  (GHz) [lin/quad/inv] &$\alpha$ (deg)& $\sigma (\alpha)$ (deg)\\
\hline
WMAP7& 53.1 / 55.4 / 47.7 & -0.9 & 2.0\\
B03& 145.0 / 147.3 / 137.8 & -4.3 & 4.1 \\
BICEP& 135.9 / 139.5 / 124.0 & -2.6 & 1.2 \\
QUAD1& 95.0 / 96.2 / 91.3 & -1.9 & 2.3\\
QUAD2& 150.0 / 151.9 / 144.3 & 0.8 & 1.1\\
\hline
\end{tabular}
\caption{For each experiment considered, we report the effective energy (for the different kinds of functional energy dependence under investigation, except the energy independent case), the corresponding measure of $\alpha$ and the associated error (statistical and systematic combined). }
\label{tab:energychannels}
\end{table}

We will treat each experiment as independent. In fact, each instrument is characterized by a different sky coverage and instrumental noise properties. 
Confirmation of the fact that the dataset can be safely assumed as independent comes from the agreement between the two combined analysis of \cite{Xia:2012ck} and \cite{Komatsu:2010fb}. The former performs a joint analysis of all the angular power spectra, while the latter only combines the final results derived from each experiment.

Note that for QUAD, that provided a separate analysis of the two frequency channels, we consider the two  data independently, to improve tests of energy dependence. Moreover this allows us to study the influence of the $150 \text{GHz}$ channel data on the final  fits. As we have mentioned in the previous section, this channel provides a constraint on $\alpha$ that has a $\sim 2\sigma$ tension with the combination of all other data, both from other experiments and  from the $100 \text{GHz}$ channel of the same experiment. A detailed analysis of the power spectra arouses suspicions about the possible presence of unaccounted systematic errors  \cite{Xia:2009ah}. So we think it is worth to treat this issue in a careful way, to avoid hasty conclusions about the fate of birefringence models.

\subsection{Results}

Our best-fit parameters are reported in table \ref{tab:results}, for two cases: including (left) or not (right) the QUAD 150GHz channel in the analysis. We have fitted for energy-independent behavior ($\alpha(E)=\alpha_{0}$), linear energy dependence ($\alpha(E)=l E$), quadratic energy dependence ($\alpha(E)=q E^{2}$) and inverse quadratic dependence ($\alpha(E)=h E^{-2}$). 
Fit parameters and corresponding uncertainties have been computed as shown in Appendix \ref{sec:fiterror}.

\begin{table}[htdp]
\centering
\begin{tabular}{|c||c|c||c|c|}
\hline
&\multicolumn{2}{|c||}{\it with QUaD 150 GHz} & \multicolumn{2}{|c|}{\it without QUaD 150 GHz} \\
\hline
parameter&fit value $\pm$ error &  red. $\chi^2$&fit value $\pm$ error &  red. $\chi^2$\\
\hline
$\alpha_0\,\, [\text{deg}]$&$-0.93 \pm 0.70$&1.4&$- 2.22\pm 0.92$&0.17\\
$l \,\,\,[\text{deg}\cdot \text{GHz}^{-1}]$&$-6.1 \cdot 10^{-3} \pm 5.3\cdot 10^{-3}$&1.5&$-1.99 \cdot 10^{-2} \pm 0.78\cdot 10^{-2}$&0.03\\
$q\,\, [\text{deg}\cdot\text{GHz}^{-2}]$&$-3.6 \cdot 10^{-5} \pm 3.6 \cdot 10^{-5}$&1.6&$-1.44 \cdot 10^{-4} \pm 0.57\cdot 10^{-4}$&0.06\\
$h\,\, [\text{deg}\cdot\text{GHz}^{2}]$&$ -4.2 \cdot 10^3\pm 4.2 \cdot 10^3 $& 1.6 &$ -5.0 \cdot 10^3\pm 4.3 \cdot 10^3$&1.32\\
\hline
\end{tabular}
\caption{Fit parameters, together with their uncertainties and corresponding reduced $\chi^2$. On the left results obtained by including QUaD 150 GHz channel in the analysis and on the right results obtained not  including that channel.}
\label{tab:results}
\end{table}

\begin{figure}[htdp]
\centering
\includegraphics[scale=0.83]{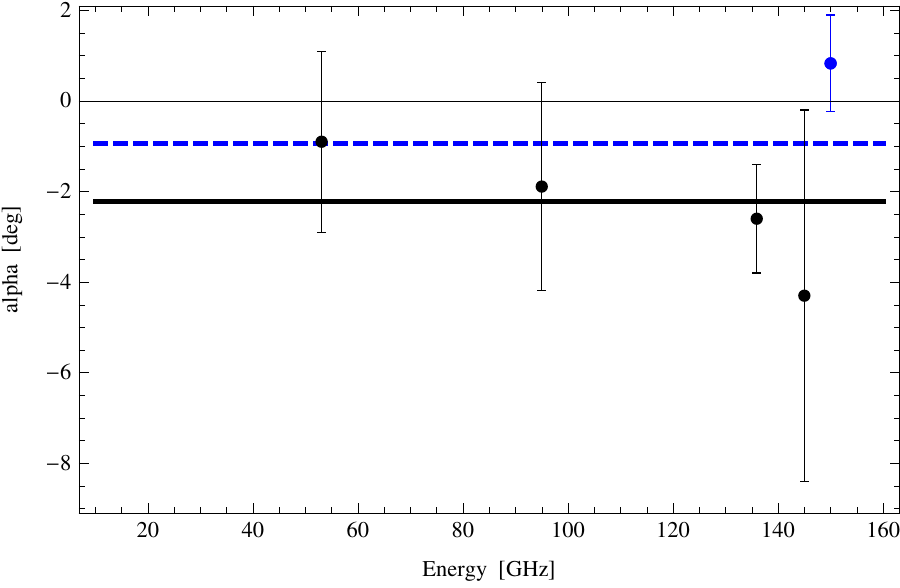}
\includegraphics[scale=0.83]{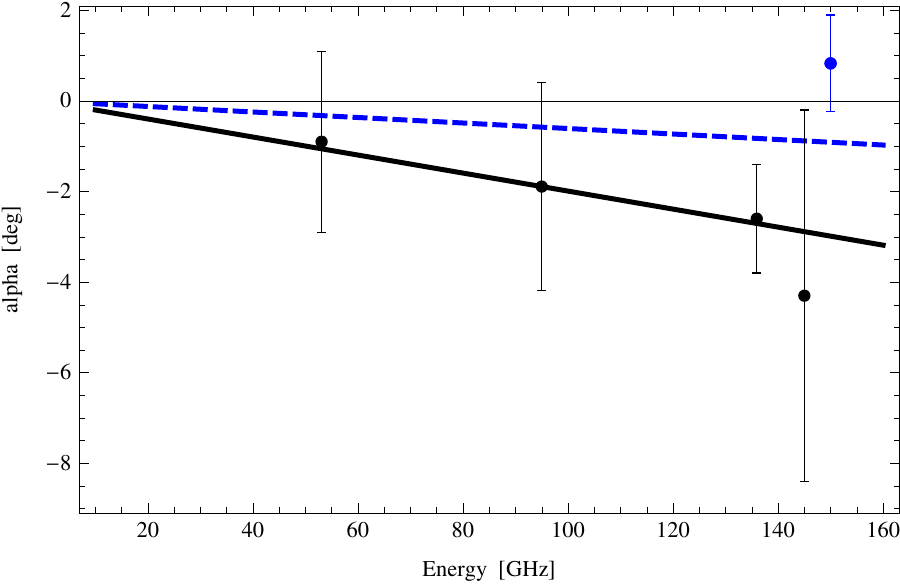}
\caption{Left panel: no energy dependence. Right panel: linear energy dependence. In both figures the thin black line marks the zero, the (dashed) blue line is the best fit curve obtained including the QUAD $150$ GHz channel (blue point), while the thick black line is the one not including the QUAD $150$ GHz channel.}
\label{fig:fit_nolin}
\end{figure}

\begin{figure}[htdp]
\centering
\includegraphics[scale=0.83]{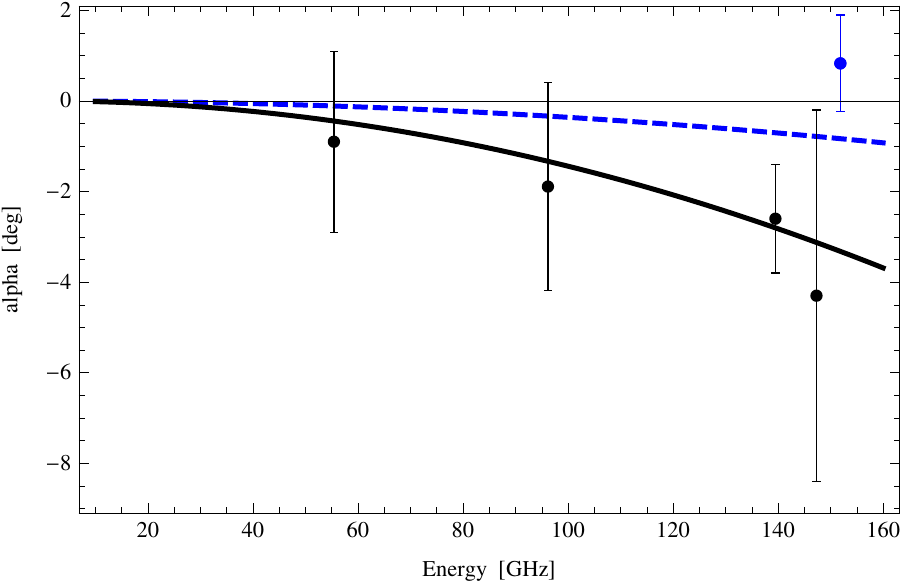}
\includegraphics[scale=0.83]{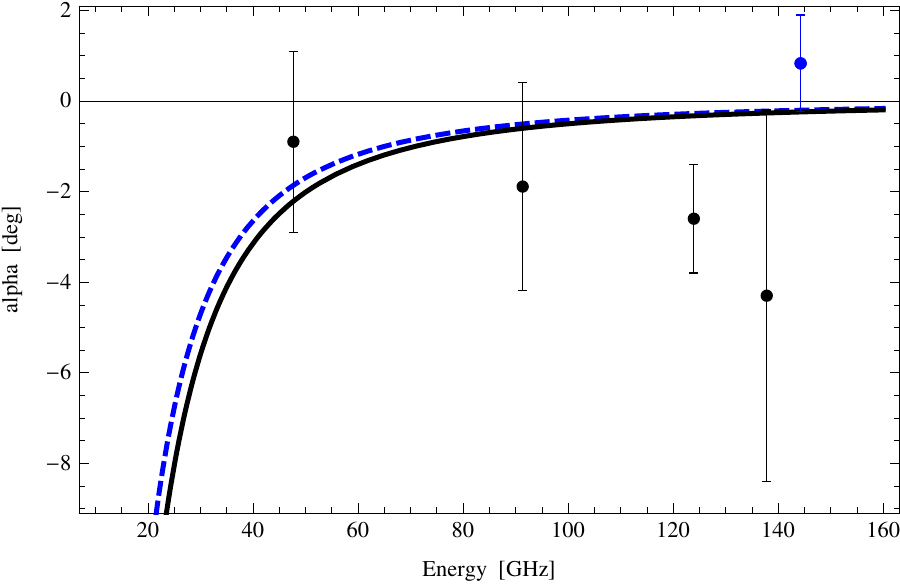}
\caption{Left panel: quadratic energy dependence. Right panel: inverse quadratic dependence. For details see caption of figure \ref{fig:fit_nolin}.}
\label{fig:fit_quadinv}
\end{figure}

Our best-fit results are also shown in Figs. \ref{fig:fit_nolin} and \ref{fig:fit_quadinv}, together with data points and associated errors (taken from table \ref{tab:energychannels}). Two curves are shown in each panel: in (dashed) blue the best-fit curve obtained by including the QUAD 150 GHz channel (blue point) in the dataset; in (solid) black, the best-fit curve obtained after rejecting the QUAD 150 GHz channel.
The black thin line is for the reference value $\alpha\equiv 0$.

We see that in general, as expected, excluding the QUAD $150$ GHz channel gives constraints on the parameters which are more significantly different from zero. In this case, while the linear and quadratic energy dependence parameters are both at a bit more than $2.5 \sigma$ from zero, the energy-independent case parameter is at $2.4 \sigma$ and the inverse energy dependence parameter is at only $1.1 \sigma$ from zero.
The $\chi^{2}$ values are all very small, except for the inverse quadratic case. So, while the inverse quadratic case is disfavored by data, we can not discriminate among the other three models, since error bars are too large.

 Nevertheless, looking at the plots we see that indeed, if one considers only the black points, they  appear to follow very well both the linear and quadratic dependence black curves (better than the other kinds of trend), but their error bars are too big to make any serious claim. We would only take this observation as an incentive to look at energy dependence of future experiments data, which are expected to provide a single-channel sensitivity on $\alpha$  about one order of magnitude better (see next section). In that case, it will hopefully be possible to make claims about the trend of the polarization rotation with respect to energy.

Including the QUAD $150$GHz channel in the analysis gives on the other hand reduced $\chi^{2}$ values around $1.5$. In this case, data disfavor all of the four considered models.
Moreover, each of the estimated fit parameters is well consistent with no effect, deviating slightly more than $1\sigma$ from zero for the constant and linear case, $1\sigma$ for the quadratic and inverse quadratic case.

We can also estimate the probability that available data are compatible with no effect at all, regardless of a possible energy dependence. To this aim we sum the residuals of each data point from the zero line ($\chi_{0}^{2}$) and provide the corresponding probability.
Including all QUAD points we obtain $\chi_{0}^{2}=7.27$ (probability $20\%$), while excluding the $150$GHz channel gives $\chi_{0}^{2}=6.67$ (probability $15\%$).
In both cases data disfavor the hypothesis that what we see is only due to fluctuations around zero, but we can not exclude it yet.

In the conclusions we will translate our findings into constraints on parameters of the models discussed at the beginning of this section. 
Here we just want to stress how crucial it seems to understand the reliability of the QUAD $150$GHz channel one one hand, and on the other to improve the sensitivity to birefringence for the single energy channels of future experiments. In fact it seems that the possibility to significantly discriminate between different models, giving stringent constraints on their parameters, is very close.

\section{Planck Forecast}

 In this section we forecast the improvement on constraining the energy dependence of $\alpha$ from the Planck mission \cite{PlanckBlue}.
We focus on the central frequency bands explored by the Planck satellite, 70GHz, 100GHz and 143GHz, whose frequency bandwidths are, respectively, $14$GHz, $33$GHz and  $48$GHz. We refer to \cite{Gubitosi:2009eu} for the sensitivity to the birefringence angle for each channel. We follow the same procedure adopted for the other CMB experiments to account for the observational bandwidth,  keeping the three energy channels separated in order to have a wide coverage in the frequency domain. The error bars associated to the best fit parameters are estimated exactly in the same way as we have done above (see Appendix \ref{sec:fiterror}), and compared to the previous results.

\begin{table}[htdp]
\centering
\begin{tabular}{|c|c|c|}
\hline
Planck channel (GHz) & Effective Energy  (GHz) [lin/quad/inv] & $\sigma (\alpha)$ (deg)\\
\hline
70 &   70.0 /  70.5 / 68.6  & 0.64\\
100&  100.0 / 101.8 / 94.4  & 0.14 \\
143& 143.0 / 145.6 / 135.1 & 0.073 \\
\hline
\end{tabular}
\caption{For each Planck channel considered, we report the effective energy (for the different kinds of energy dependence under investigation, except of course for the energy independent case), and the corresponding sensitivity on $\alpha$. }
\label{tab:planckvalues}
\end{table}

In table (\ref{tab:fitparamerrors}) we report both the current sensitivity on fit parameters and the forecasted sensitivity for Planck.
We see that using Planck data by themselves will allow to improve all constraints of about one order of magnitude. The availability of several distinct energy channels will also allow to use energy dependence to discriminate between the different energy dependence possibilities if any kind of effect is found. 

\begin{table}[h]
\centering
\begin{tabular}{|c|c|c|}
\hline
parameter & Current sensitivity & Planck sensitivity\\
\hline
$\sigma(\alpha_0)\,\, [\text{deg}]$& $7.0\cdot 10^{-1}$ & $6.4 \cdot 10^{-2}$\\
$\sigma(l)\,\,\,[\text{deg}\cdot \text{GHz}^{-1}]$ & $5.3\cdot 10^{-3}$ & $4.8\cdot 10^{-4}$ \\
$\sigma (q)\,\, [\text{deg}\cdot\text{GHz}^{-2}]$& $3.6\cdot 10^{-5}$& $3.3\cdot 10^{-6}$ \\
$\sigma(h)\,\, [\text{deg}\cdot\text{GHz}^{2}]$&$4.2\cdot 10^{3}$ &$8.7 \cdot 10^{2}$\\
\hline
\end{tabular}
\caption{Planck forecasts on fit parameter errors. Current sensitivities are based on the full dataset of table \ref{tab:energychannels}. Planck sensitivities are based on Planck  instrumental properties as reported in table \ref{tab:planckvalues}. }
\label{tab:fitparamerrors}
\end{table}

 In order to show explicitly how much improvement  on distinguishing among models is  expected for Planck, we perform the following test. We assume that the birefringence angle has a linear dependence on energy, defined by our best-fit value $l=1.99\cdot 10^{-2}$ $\text{deg}\cdot \text{GHz}^{-1}$. Then we simulate 10000 datasets,  extracting for each channel of Planck and of the other experiments we have considered in the previous section a value of $\alpha$ from a gaussian distributions centered on the model and of standard deviation defined by tables \ref{tab:energychannels} and \ref{tab:planckvalues}. We fit then each simulated dataset with a linear, constant, and quadratic dependence and store the corresponding $\chi^2$ values.  
\begin{figure}[h]
\centering
\includegraphics[scale=0.83]{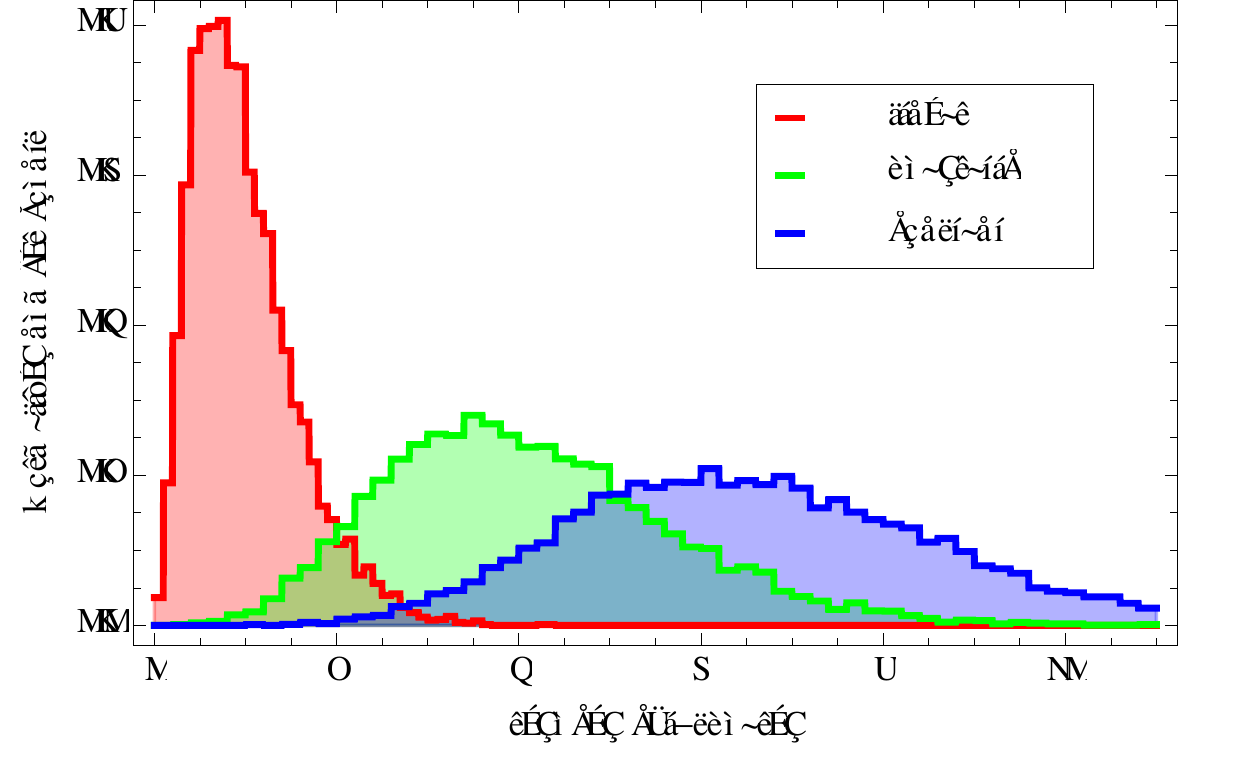}
\caption{Reduced $\chi^2$ probability distributions for the linear (red curve), quadratic (green curve) and constant (blue curve) energy dependence fitting. The 10000 simulations have been drawn from the linear model defined by $l=1.99\cdot 10^{-2}$ deg $\cdot$ GHz$^{-1}$.}
\label{fig:chi2_comp}
\end{figure}
In Fig. \ref{fig:chi2_comp} we report the reduced $\chi^2$ distributions for the three cases. Whereas the linear case (red curve) is very well peaked around 1 (consistently with the model we assumed), the quadratic (green curve) and constant (blue curve) cases are, almost always, very poorly fitted. In particular, fitting for a quadratic dependence, the probability for the $\chi^2$ value to be lower than as fitting for the linear case is as low as $1.53\%$. For a constant dependence, such a value further drops to $0.06\%$.

\section{Discussion and conclusions}

In this work we have fitted  available data on birefringence coming from CMB polarization measurements against several possible relations between the birefringence effect and photons energy.
This is the first time that the issue of energy dependence is treated in such a systematic way. 

In principle this kind of analysis would allow to disentangle effects of different origin that result in the same kind of rotation of the polarization direction (like Chern-Simons interactions of the EM field, propagation of CMB photons through primordial magnetic fields, Planck scale Lorentz violations just to mention a few). 
Even if polarization data are not good enough to prefer a specific trend among the competing ones investigated in this work, we are able nonetheless to improve constraints on the underlying theoretical models up to one order of magnitude with respect to previous results.

For example, we have studied the possibility of a quadratic energy dependence of the birefringence effect. This might be traced back to Quantum Gravity Planck-scale effects \cite{Myers:2003fd, Gubitosi:2009eu},  whose relics at low energies can be modeled as a coupling of the EM field with a fixed time-like vector.
If, following \cite{Myers:2003fd} and \cite{Gubitosi:2009eu}, we write the coupling constant between the EM field and the vector through a dimensionless parameter $\xi$ and the Planck mass scale $M_{P}$, then the polarization rotation angle is related to the distance covered by photons and their energy by:
\begin{equation}
\alpha(E)=\frac{\xi}{M_{P}}E^{2}\Delta\ell \,.
\end{equation}
Taking into account photon energy redshift from the last scattering surface ($z_{LS}$) this becomes
$
\alpha(E_{0})=\frac{\xi}{M_{P}}E_{0}^{2}\int_{0}^{z_{LS}}(1+z)H(z)^{-1}dz
$, where $E_{0}$ is the photon energy today.
So our constraint $q=(-3.6\pm3.6)\cdot 10^{-5}\, \text{deg}\cdot \text{GHz}^{-2}$ can be translated\footnote{Here and in the following we assume a cosmological model with $H_{0}\equiv h_{0} \cdot 100\, \text{Km/s/Mpc}=71   \,\text{Km/s/Mpc}$, $\Omega_{m}=0.27$ and $\Omega_{\Lambda}=0.73$.} into a constraint on the Quantum Gravity  parameter $\xi=- 0.22\pm0.22$. If we disregard the QUAD $150$GHz channel, the constraint  $q=(-1.44\pm0.57)\cdot 10^{-4}\, \text{deg}\cdot \text{GHz}^{-2}$ translates into $\xi=-0.87 \pm 0.34$. To compare with previous constraints from CMB \cite{Gubitosi:2009eu}, $\xi'=-0.110\pm0.075$, we have to take into account the different convention used in treating the Planck mass scale. In \cite{Gubitosi:2009eu} it was used the reduced Planck mass $M_{P}^{(r)}\equiv M_{P}/\sqrt{8\pi}$. 
We see in this way that our constraint improves the previous one of about a factor three.

Concerning the linear energy dependence case, our results can be translated into constraints on the `Weyl' interaction described in \cite{Shore:2004sh} (see also \cite{Kahniashvili:2008va}). 
We are not giving here the details of the model, which is well discussed in the two references given above. We just report that, in this case, the polarization rotation angle is related to the distance $\Delta \ell$ traveled by photons and the dimensionless scalar $\Psi_{0}$, which  sets the amplitude of the interaction, by  
\begin{equation}
|\alpha(\omega)|=4\omega|\Psi_{0}|\Delta \ell \,.
\end{equation}
Taking into account the Universe expansion and the energy redshift,  our parameter  for the linear dependence is related to $\Psi_{0}$ through $|l|=8\pi|\Psi_{0}|\int_{0}^{z_{LS}}H(z)^{-1}dz$.
Assuming gaussian errors for our parameter $l$, so that the limit on its absolute value is $|l|\leq 8.7\cdot 10^{-3}\,\text{deg/GHz}$ at 68\% c.l.  ($|l|\leq 1.5 \cdot 10^{-2}\,\text{deg/GHz}$ at 95\% c.l.), we get  $|\Psi_{0}|\leq 5.8 \cdot 10^{-33}h_{0}$ at 68\% c.l. ($|\Psi_{0}|\leq 1.0 \cdot 10^{-32}h_{0}$ at 95\% c.l. ), where $h_{0}$ is the reduced Hubble constant \footnote{Note that this kind of `Weyl' interaction would also produce a leakage from linear to circular polarization, which could be detectable by looking for a circular component of the CMB.}. 
Such a constraint improves the previous one \cite{Kahniashvili:2008va} of about one order of magnitude. 
Disregarding the QUAD $150$GHz channel, our $68$\% c.l. upper limit on $l$ is $|l|\leq 2.4 \cdot 10^{-2}\,\text{deg/GHz}$ ($|l|\leq 3.3\cdot 10^{-2}\,\text{deg/GHz}$ at 95\% c.l.), so  $|\Psi_{0}|\leq 1.6\cdot 10^{-32}\, h_{0}$ at $68$\% c.l. ($|\Psi_{0}|\leq 2.2 \cdot 10^{-32}\, h_{0}$ at $95$\% c.l.)

The constraint on the inverse quadratic energy dependence of the birefringence effect can be translated into a constraint on the amplitude of uniform primordial magnetic fields. 
Using the result of \cite{Kosowsky:1996yc} and \cite{Campanelli:2004pm},  we can relate the \textit{rms} value of the rotation angle to the intensity of the primordial magnetic field through: 
\begin{equation}
<\alpha^{2}>^{1/2}=1.3^{\circ}\left(\frac{B_{0}}{10^{-9}\text{Gauss}}\right)\left(\frac{\nu_{0}}{30\text{GHz}}\right)^{-2} \,.
\end{equation}
So our parameter $h$ is related to the magnetic field intensity in the following way: $<h^{2}>^{1/2}=1.3^{\circ}\left(\frac{B_{0}}{10^{-9}\text{Gauss}}\right)\left(30\text{GHz}\right)^{2}$. From our constraint on $h$, assuming gaussian errors we derive the upper limit $<h^{2}>^{1/2}< 6.3\cdot 10^{3}$ at 68\% c.l., which translates into the constraint on the magnetic field intensity $B_{0}\leq 5.4 \cdot10^{-9}\,\text{Gauss}$  and $B_{0}\leq 6.1 \cdot10^{-9}\,\text{Gauss}$ at 68\% c.l., with and without the QUAD $150$GHz channel respectively.
Of course this result is valid only under the assumption of a uniform primordial magnetic field. In most general scenarios the magnetic field will have some non trivial spectral distribution, which would result in a direction-dependent birefringence (see e.g. \cite{Pogosian:2011qv}).

As last case, we consider  energy independent birefringence, which is the  most broadly studied in the literature. We can translate our constraint $\alpha_{0}=(-0.93\pm0.70)^{\circ}$, obtained including the QUAD 150GHz channel in the analysis, into an upper limit $|\alpha|<1.26^{\circ}$ at $68$\% c.l. and $|\alpha|<2.11^{\circ}$ at $95$\% c.l.  This can be used to constrain the Chern-Simons interaction of the EM field described in \cite{Carroll:1989vb}. The relevant parameter of the model is $|L_{0}-L\cos \phi|$, which sets the intensity of the interaction. The birefringence angle is related to this parameter and the distance covered by photons ($\Delta\ell$) by 
\begin{equation}
|\alpha|=\frac{1}{2}|L_{0}-L\cos\phi|\Delta \ell \,.  \footnote{ $\phi$ is the angle between a preferred direction pointed out by the vector $L_{\mu}$ and the photon propagation direction. Since we use data averaged out over the whole sky, the  limits we put  have to be interpreted as limits on the average value of $|L_{0}-L\cos\phi|$ over all directions. }
\end{equation}
 Taking $\Delta\ell\equiv \int_{0}^{z_{LS}} (1+z)^{-1}H(z)^{-1}dz=2.15 \cdot 10^{-33}  h_{0}\,\text{eV}$ we find $|L_{0}-L\cos\phi|<9.4 \cdot 10^{-44} h_{0}\, \text{GeV}$ at $68$\% c.l. and $|L_{0}-L\cos\phi|<1.6 \cdot 10^{-43} h_{0} \,\text{GeV}$ at $95$\% c.l. This constraint is slightly stronger than the one found in \cite{Kahniashvili:2008va}, since we are taking into account a larger dataset.
When disregarding the $150$GHz QUAD channel, our constraint $\alpha_{0}=(-2.22\pm0.92)^{\circ}$ translates into the upper limit  $|\alpha_{0}|<2.66^{\circ}$ at $68$\% c.l., and $|\alpha_{0}|<3.77^{\circ}$ at $95$\% c.l. In this case $|L_{0}-L\cos\phi|<2.0\cdot 10^{-43} h_{0}\, \text{GeV}$ at $68$\% c.l. and $|L_{0}-L\cos\phi|<2.8 \cdot 10^{-43} h_{0} \,\text{GeV}$ at $95$\% c.l. 
Note that  limits on the constant energy dependence have to be taken with a grain of salt, since the same kind of polarization rotation might also be mimicked by systematic errors in polarimeters calibration \cite{systematics,Yadav:2012tn}.

\section*{Acknowledgements}
We thank Davide Franco and Ken Ganga for useful discussions during the development of this work. We also thank Fabio Finelli, Matteo Galaverni and Giovanni Amelino-Camelia for comments on the manuscript.

\appendix

\section{Multi-channel experiments}
\label{sec:effectiveEnergy}
In this appendix we provide explicit calculations of the effective energy for a multi-channel experiment. The four energy dependence of $\alpha$ under investigation are considered.

\begin{itemize}
\item{No dependence ($\alpha (E)=\alpha_{0}$)}. This case is trivial, since $\alpha_{f}=\alpha_{0}$ and no effective energy is introduced. 

\item{Linear dependence ($\alpha(E)=l E$)}.
\begin{equation}
\alpha_f=\frac{\int l E f(E)dE}{\int f(E)dE}  \equiv l E_f \Rightarrow E_f=\left[\frac{1}{2}\frac{\sum_{i}\frac{1}{\sigma_{i}^{2}}\left((E_+^{\,(i)})^2-(E_-^{\,(i)})^2\right)}{\sum_{i}\frac{1}{\sigma_{i}^{2}}\left(E_+^{\,(i)}-E_-^{\,(i)}\right)}\right]\,.
\end{equation}

\item{Quadratic dependence ($\alpha(E)=q E^{2}$)}. 
\begin{equation}
\alpha_f=   \frac{\int q E^{2} f(E)dE}{\int f(E)dE} \equiv q E^2_f \,,\Rightarrow E_f= \left[ \frac{1}{3} \frac{\sum_{i}\frac{1}{\sigma_{i}^{2}}\left((E_+^{\,(i)})^3-(E_-^{\,(i)})^3\right)}{\sum_{i}\frac{1}{\sigma_{i}^{2}}\left(E_+^{\,(i)}-E_-^{\,(i)}\right)} \right]^{1/2}\,.
\end{equation}

\item{Inverse square law ($\alpha(E)=h E^{-2}$)}. 
\begin{equation}
\alpha_f=   \frac{\int h E^{-2} f(E)dE}{\int f(E)dE} \equiv h E^{-2}_f\Rightarrow E_{f}=\left[\frac{\sum_{i}\frac{1}{\sigma_{i}^{2}}\left((E_-^{\,(i)})^{-1}-(E_{+}^{\,(i)})^{-1}\right)}{\sum_{i}\frac{1}{\sigma_{i}^{2}}\left(E_+^{\,(i)}-E_-^{\,(i)}\right)}\right]^{-1/2}\,.
\end{equation}
\end{itemize}

For WMAP, the pixel noise level of the  channels we consider is \cite{lambdaWMAP}, $\sigma_1=2.672 \,\text{mK}$ for the Q-band ($41$ GHz), $\sigma_{2}=3.371\, \text{mK}$ for the V-band ($61$ GHz) and $\sigma_{3}=6.877\,  \text{mK}$ for the W-band ($94$ GHz), while for BICEP $\sigma_{1}=0.81 \,\mu\text{K}$ for the $96$ GHz channel and $\sigma_{2}=0.64 \, \mu\text{K}$ for the $150$ GHz channel  \cite{Chiang:2009xsa}.

\section{Fitting procedure}
\label{sec:fiterror}
Let $y_i$ be our $i$-th measure (of $\alpha$) at the energy $x_i$, and $\sigma_i$ its associated error. Assuming a functional form $f(x,p)$ as description of the data, the fit parameter $p$ is found by minimizing the $\chi^2$:
\begin{equation}
\frac{\partial \chi^2}{\partial p}=\frac{\partial }{\partial p} \sum_i\frac{\left( y_i-f(x_i,p)\right)^2 }{\sigma_i^2}\qquad \equiv 0\,.
\end{equation} 
The uncertainty associated to the fit parameter can then be computed as 
\begin{equation}
\sigma^2(p)=\sum_i\left( \frac{\partial p}{\partial y_i}\right)^2 \sigma_i^2\,.
\end{equation}

Here below, we provide the analytic expression for the fit parameters and associated uncertainties for the four energy dependences considered in the present work.

\begin{itemize}

\item{No dependence ($\alpha (E)=\alpha_{0}$)}. 
\begin{equation}
\alpha_0= \left(\sum_i \frac{y_i}{\sigma^2_i}\right) \left(\sum_i \frac{1}{\sigma^2_i}\right)^{-1} ,\qquad \qquad \sigma(\alpha_0)=\left(\sum_i \frac{1}{\sigma_i^2}\right)^{-1/2}. 
\end{equation}

\item{Linear dependence ($\alpha(E)=l E$)}.
\begin{equation}
l= \left(\sum_i \frac{x_i y_i}{\sigma^2_i}\right) \left(\sum_i \frac{x_i^2}{\sigma^2_i}\right)^{-1}, \qquad \qquad \sigma (l)=\left(\sum_i \frac{x_i^2}{\sigma_i^2}\right)^{-1/2} .
\end{equation}

\item{Quadratic dependence ($\alpha(E)=q E^{2}$)}. 
\begin{equation}
q= \left(\sum_i \frac{x_i^2 y_i}{\sigma^2_i}\right) \left(\sum_i \frac{x_i^4}{\sigma^2_i}\right)^{-1} ,\qquad \qquad \sigma (q)=\left(\sum_i \frac{x_i^4}{\sigma_i^2}\right)^{-1/2}. 
\end{equation}

\item{Inverse square law ($\alpha(E)=h E^{-2}$)}. 
\begin{equation}
h= \left(\sum_i \frac{y_i}{x_i^2 \sigma^2_i}\right) \left(\sum_i \frac{1}{x_i^4\sigma^2_i}\right)^{-1} ,\qquad \qquad \sigma (h)=\left(\sum_i \frac{1}{x_i^4 \sigma_i^2}\right)^{-1/2}.
\end{equation}
\end{itemize}

\bibliographystyle{apsrev4-1}
\bibliography{bibliography}

\end{document}